\documentclass[aps,prl,reprint,longbibliography]{revtex4-2}
\usepackage{amsmath,amssymb} 
\usepackage{amsmath} 
\usepackage{amsfonts} 
\usepackage{bm} 
\usepackage[dvipdfmx,colorlinks=true,citecolor=blue,linkcolor=blue,filecolor=blue,urlcolor=blue]{hyperref} 
\usepackage[dvipdfmx]{graphicx} 
\usepackage{braket} 

\begin{document}

\title{Thermal Pure Quantum Matrix Product States \textcolor{black}{Recovering a Volume Law Entanglement}}

\author{Atsushi Iwaki}
\email{iwaki-atsushi413@g.ecc.u-tokyo.ac.jp}
\author{Akira Shimizu$^\star$}
\author{Chisa Hotta}
\email{chisa@phys.c.u-tokyo.ac.jp}
\affiliation{Department of Basic Science, University of Tokyo, Meguro-ku, Tokyo 153-8902, Japan}
\affiliation{$\:^\star$ Komaba Institute for Science, The University of Tokyo, Meguro-ku, Tokyo 153-8902, Japan}
\date{\today}

\begin{abstract} 
We propose a way to construct thermal pure quantum matrix product state (TPQ-MPS) 
that can simulate finite temperature quantum many body systems with a minimal numerical cost comparable 
to the matrix product algorithm for the ground state. 
\textcolor{black}{
The MPS was originally designed for the wave function with area-law entanglement. 
However, by attaching the auxiliary sites to the edges of the random matrix product state, 
we find that the degree of entanglement is automatically tuned so as to recover the volume law of the entanglement entropy 
that characterizes the TPQ state. 
}
The finite temperature physical quantities of the transverse Ising and the spin-1/2 Heisenberg chains 
evaluated by a TPQ-MPS show excellent agreement even for bond dimension $\sim 10$-$20$ 
with those of the exact results. 
\end{abstract}

\maketitle

\par
\textit{Introduction.}
Finding a good description of typical wavefunctions of quantum many body states at finite temperature 
has been a challenge in condensed matter theory. 
Traditional quantum mechanical representation of equilibrium states rely on the density operators 
of Gibbs' ensembles, 
which are the classical mixtures of an exponentially large numbers of pure quantum states. 
This construction does not allow us to simulate sufficiently large quantum systems of physical interest, 
since the available numerical devices, e.g. stochastic quantum Monte Carlo (QMC) methods or 
some size-free methods\cite{oitmaa10,bishop00,chisa18,bernu20} are limited. 
\par
The key conceptual development against the ensemble physics is 
the typicality\cite{vonneumann29,imada86,sugita07,popescu06,goldstein06,reimann07}; 
there exists {\it a single} thermal pure quantum (TPQ) state that solely represents the thermal equilibrium\cite{hams00,machida05,machida12,iitaka03,sugiura12,sugiura13,hyuga14}. 
It then happens that for the description of any of the equilibrium quantum states, 
one can choose an arbitrary degrees of classical mixture, 
from the purity-$1/(e^{\Theta(N)})$ ensemble to purity-1 TPQ state. 
Let us consider a size-$N$ system with a Hamiltonian ${\mathcal H}$, and its subsystem-A of size-$n$. 
Since the equilbrium state at inverse temperature $\beta$ 
can be described equivalently by the TPQ state $|\psi\rangle_\beta$ and by the Gibbs state, 
the local observables ${\cal O}_A$ also fulfill 
\begin{equation}
_\beta\langle \psi| {\mathcal O_A} |\psi\rangle_\beta
={\rm tr} ({\mathcal O_A} e^{-\beta {\mathcal H}})/{\rm tr} (e^{-\beta {\mathcal H}}). 
\end{equation}
This directly indicates the equality of the local density matrix and the local canonical ensemble, 
$\rho_n( |\psi\rangle_\beta )={\rm tr}_{\bar A} ( e^{-\beta {\mathcal H}})/{\rm tr} (e^{-\beta {\mathcal H}})$. 
Accordingly, the von Neumann entropy $S_n=-{\rm Tr}(\rho_n\ln \rho_n)$ 
obtained by the reduced density operator $\rho_n$ of the subsystem-A 
becomes the entanglement entropy(EE) of a TPQ state, 
and is related to thermodynamic entropy density $s_{th}$ as \cite{eq2proof}
\begin{equation}
S_n/n = s_{th}, (1\ll n\ll N). 
\label{equality}
\end{equation}
The EE of the TPQ state thus needs to fulfill the volume law, and indeed, 
in the similar context, the Page curve of the second Renyi entropy in 
a finite open boundary system is observed in the exact TPQ-state\cite{nakagawa18}. 
\par
Practically, however, constructing such exact TPQ state, which we call full-TPQ state\cite{hams00,sugiura12,sugiura13}, 
requires a cost only slightly smaller than the conventional finite temperature diagonalization methods\cite{jaklic94,aichhorn03,shnack20}, and is available only up to $\sim 2N$ of that of the Gibbs state. 
As for the ground state, the approximate forms of the pure wave functions are established by 
the density matrix renormalization(DMRG) or matrix product states (MPS)\cite{fannes89,fannes91,verstraete06}. 
However, their application to the TPQ state has never been tested because 
the MPS description has an area law entanglement by construction\cite{amico08,eisert10}, 
and is apparently unsuitable for the finite temperature case where the entanglement blows up massively. 
\par
In this Letter, we show that the TPQ-MPS state is realized 
by attaching appropriate auxiliary sites at both edges of the system, 
which work as an entanglement-bath and make the system highly entangled. 
By successively operating the Hamiltonian to the random matrix product state (RMPS) with the auxiliaries, 
the MPS is annealed down to lower temperature 
\textcolor{black}{
where we find that the system recovers the volume law entanglement 
when measured from the system center toward the very edges. 
}
%
\begin{figure*}
   \centering
   \includegraphics[width=17.5cm]{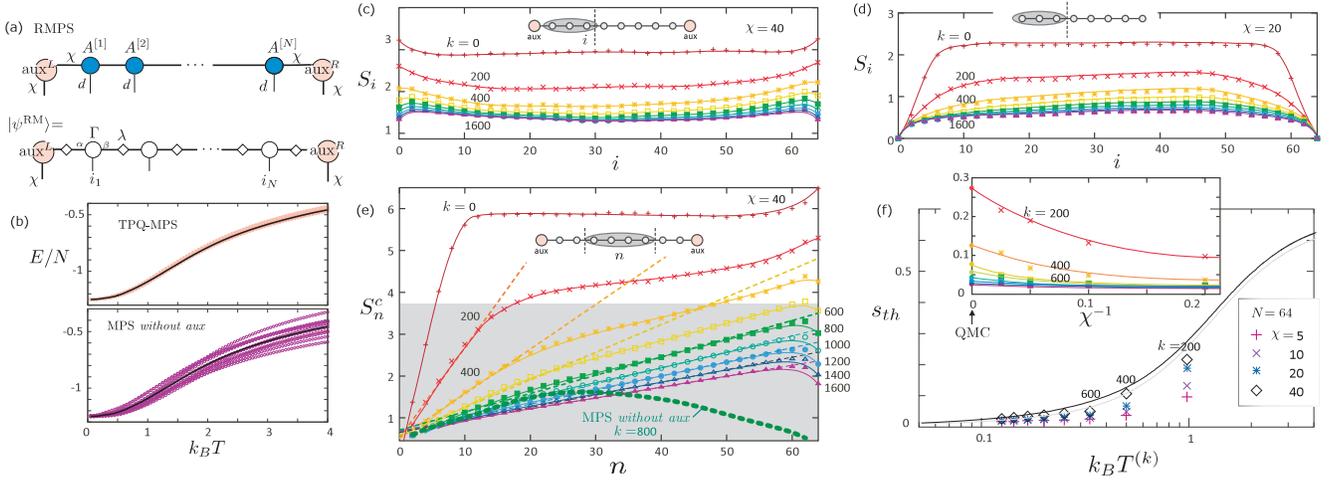}
   \caption{(a) Schematic representation of the RMPS state in Eq.(\ref{rmps}). 
  Two auxiliary sites (Red circles) are connected to the two open edges of the system, 
  which has no physical interactions with the main system. 
   Lower panel is the canonical form of the TPQ-MPS $|\psi^{\rm RM}\rangle$. 
   For the starting point of mTPQ ($k=0$) 
   we take $\Gamma^{[m]i_m}_{\alpha_m\beta_m}$ of $\chi\!\times\!\chi\!\times\!d$ with 
   a Gaussian distribution, $\lambda_i=1$, and ${\rm aux}^{X}_{\alpha_{X} \beta}$ 
   of $X=L, R$ as the $\chi\!\times\!\chi$ unit matrices 
   which gives the canonical form of RMPS. 
   \textcolor{black}{
  (b) Energy density of the transverse Ising model plotted independently for 10 mTPQ-runs. Upper/lower panels 
     are those of TPQ-MPS and MPS without auxiliaries with $N=16$ and $\chi=40$. 
  (c,d) Entanglement entropy $S_i=-{\rm Tr}(\hat \rho_i \ln \hat \rho_i)$ 
    of the TPQ-MPS and MPS without auxiliaries, when the $N=64$ system is divided at bond $i=1\sim N$. 
    The $k$-th mTPQ state of the transverse Ising model with fixed $\chi=40,20$ are shown.  
  (e) Entanglement entropy $S_n^c=-{\rm Tr}(\hat \rho_n \ln \hat \rho_n)$ 
    of the TPQ-MPS when dividing $N=64$ system and picking up the $n$-sites from the center.
    Broken lines are the linear fits whose slope gives the thermodynamic entropy $s_{th}$. 
    Bold dotted line is $S_n^c$ obtained without auxiliaries for $k=800$. 
    The shade marks $S_n^{c}\lesssim \ln\chi$ allowed for the standard MPS for $\chi=40$. 
  (f) Slope of $S_n^c$ as a function of $k_BT^{(k)}$ for $k=200$-$1600$ in the mTPQ calculation 
    for $\chi=5,10,20,40$ and $N=64$. Solid and gray lines are $s_{th}$ obtained by the QMC results at $N=64$ and 32. 
    Inset: $\chi^{-1}$ dependence of the data where the circles at the starting point are the QMC results for the corresponding temperature. 
 }
   }
\label{fig1}
\end{figure*}
We demonstrate that the TPQ-MPS wave functions give accurate evaluation of the physical quantities 
in typical quantum spin models {\it without taking ensemble average}. 
The computational cost is significantly reduced to that of the MPS of the ground state, 
and the accessible system size increases to $N \gtrsim 100$. 
\par
So far, the MPS methods at finite temperature like minimally entangled typical thermal states (METTS)
\cite{white09,stoudenmire10} and matrix product purification(MPP)\cite{verstraete04, zvolak2004, feiguin05}, 
both with intermediate degrees of purity, have relied on some sort of ``ensemble" averages to compensate 
for the low entanglement properties of MPS. 
The METTS starts from the minimally entangled product state which requires 
a large number of sampling average of $\gtrsim 100$. 
The MPP adopts extra $N$-ancillary system which are traced out to obtain the mixed state of the system\cite{umezawa82}; 
This expression is mathematically redundant since they replace a mixed state in 
an irreducible representation in the minimum Hilbert space of dimension $d^N$ by a reducible representation 
in a space of dimension $d^{2N}$, and does not save our numerical cost by itself. 
\textcolor{black}{
In our TPQ-MPS construction, the redundancy is limited to  $d^{N + \chi}/d^N = d^\chi$, 
which does not grow with increasing $N$\cite{comparison}. }
\par
\textit{Random initial state.} 
We consider a one-dimensional(1D) lattice system consisting of $N$ sites 
with open boundary condition (OBC) 
where each site hosts $d$-dimensional degrees of freedom, 
and two auxiliaries attached at both edges having $\chi$-dimensions. 
The RMPS of such a system shown in Fig.\ref{fig1}(a) is given as\cite{garnerone10} 
\begin{eqnarray}
|\psi^{\rm RM}\rangle &=& \sum_{\alpha _L \{ i_n\} \alpha _R} \langle {\rm aux}^L_{\alpha_L} | 
A^{[1]i_1}\cdots A^{[N]i_N} | {\rm aux}^R_{\alpha_R} \rangle 
\nonumber\\
&& \vspace{-5mm} \rule{2cm}{0mm} |\alpha_L, i_1\cdots i_N, \alpha_R \rangle , 
\label{rmps}
\end{eqnarray}
where the $A^{[m]i_m}$ is the $d\chi^2$ matrix on the $m$-th site with $i_m=1,\cdots,d$, 
and is explicitly given using the $d \chi\times d\chi$ random unitary matrix $U$ as 
$A_{\alpha\beta}^{[m]i_m}= U_{(i,\alpha)(1,\beta)}$ or $U_{(1,\alpha)(i,\beta)}$ which 
fulfill the left or right canonical form, respectively. 
Here $|{\rm aux}^{L/R}_{\alpha_{L/R}} \rangle $ is the right/left auxiliary state with $\alpha_{L/R}=1$-$\chi$.  
This RMPS reproduces the physical quantities at $T=\infty$ with the variance of order $\chi^{-2}$, 
which can be shown analytically as follows; 
taking $A$ on the l.h.s/r.h.s. 
of a one-site operator $\hat {\mathcal O}_i$ as left/right canonical form, 
we have $\langle \psi^{\rm RM}|\psi^{\rm RM}\rangle =\chi$. 
By taking account of the formula for the random average, 
$\overline{U_{ij}U_{kl}^*}=\delta_{ij}\delta_{kl}/(\chi d)$ and 
$\overline{U_{ij}U^*_{ik}U_{lm}U_{ln}^*}
=\big(\delta_{jk}\delta_{mn}+\delta_{il}\delta_{jn}\delta_{mk}
 -(\delta_{il}\delta_{jk}\delta_{mn}+\delta_{jn}\delta_{mk})/(\chi d) \big)/(\chi^2d^2-1)$, 
we have the expectation values, $\langle \psi^{\rm RM} | \cdots |\psi^{\rm RM} \rangle / \chi \equiv \langle \cdots \rangle$ as 
\begin{eqnarray}
\overline{\langle  \hat{\mathcal O}_i \rangle} 
&=& \sum_{i,j,\alpha,\beta} \frac{1}{\chi^2 d} \delta_{ij} \langle j|\hat{\mathcal O}_i|i\rangle 
= \langle {\mathcal O}_i \rangle_\infty
\label{randomav}
\end{eqnarray}
with $\langle {\mathcal O}_i \rangle_\infty\equiv d^{-1} {\rm Tr} {\mathcal O}_i$, 
and its variance as 
\begin{eqnarray}
\overline{\langle{\mathcal O}\rangle ^2 }- ( \overline{\langle{\mathcal O}\rangle} ) ^2
= \frac{d-1}{\chi^2d^2-1}  \big(
\langle{\mathcal O}_\infty^2\rangle-\langle{\mathcal O}\rangle_\infty^2 \big). 
\label{randomvariance}
\end{eqnarray}
Typicality of the RMPS is studied and confirmed numerically in the similar context\cite{garnerone10,garnerone10-2}, 
followed by several proposals to stochastically construct microcanonical and canonical ensembles of RMPS\cite{garnerone13,garnerone13-2,iitaka2020}. 

In our work, the RMPS is constructed not by using Eq.(\ref{rmps}) 
but by preparing a tensor $\Gamma^{[m]i_m}_{\alpha_m\beta_m}$ ($\in {\mathbb C}$) of bond dimension $\chi$, 
whose elements follow the Gaussian distribution (see Fig.\ref{fig1}(a)). 
It can be straightforwardly shown that after transforming Eq.(\ref{rmps}) into the canonical form by the 
successive Schmidt-decomposition\cite{shi06,vidal03}, the obtained matrices also form an equivalent RMPS. 
\par
\textit{mTPQ-MPS Method.} 
The initial state ($k=0$) is taken as the aforementioned RMPS state 
with bond dimension $\chi$, where we take auxiliaries of $\chi\times \chi$ attached at both edges 
as a unit matrix $\hat I$ in the first place. 
The RMPS successively generates a series of unnormalized 
microcanonical TPQ (mTPQ) states $k=0,1,2,\cdots$ as\cite{sugiura12} 
\begin{equation}
 \ket{k} = (l-\Hat{h}) ^k \ket{\psi^{\rm RM}}.
\label{mtpq} 
\end{equation}
where $\Hat{h}$ is the Hamiltonian divided by $N$, 
and $l$ is a parameter larger than the maximum eigenenergy, 
which is necessary to generate sharp microcanonical energy distribution(see Ref.[\onlinecite{sugiura12}].) 
Here, $\ket{k}$ is the TPQ state at a temperature $k_BT^{(k)} = N(l-u_k)/2k$\cite{footcomment2} 
with energy $u_k=\langle k | \Hat{h}| k \rangle/ \langle k | k \rangle$. 
In our algorithm, $(l-\Hat{h})$ is represented by a matrix product operator (MPO) of 
bond-dimension $D$ that depends on $\Hat{h}$, 
and applying this MPO at each step to $\ket{k}$ multiplies the matrix dimension to $D\chi$. 
Before truncating the dimension of the enlarged matrix down to $\chi$, 
we transform the MPS to its canonical form including auxiliary sites
in order to minimizes the truncation error\cite{footcomment3}. 
The process is repeated until $k_BT^{(k)}$ reaches low enough temperature $\beta_{\rm max}^{-1}$, 
and the effective bond dimension, $\chi^{\it eff}_i (i=0, N)$, 
which is the number of finite eigenvalues of the Schmidt decomposition $\lambda_i$ on the $i$-th bond, 
can change automatically within $1\le \chi^{\it eff}_i \le \chi$. 
\par
We consider an operator $\hat A$ that can be described by a low-order polynomial of local observables. 
At each step, such $\hat A$ is evaluated and stored, 
$\langle \hat A \rangle=\langle k |\hat A |k\rangle / \langle k|k\rangle$ 
is the physical quantities at $k_BT^{(k)}$. 
Instead of directly adopting this form, 
one can generate the physical quantities for arbitrary temperatures $\beta^{-1}\gtrsim \beta_{\rm max}^{-1}$
by the canonical summation as (see Ref.\cite{sugiura13}),
$
\langle A\rangle_{\beta,N} 
=\big(e^{-\beta N l} \sum_{k} 
\frac{(\beta N/2)^{2k} }{(k!)^2}  \;\langle k | \hat A |k \rangle 
+ \frac{(\beta N/2)^{2k+1} }{(k+1)!(k!)} \;\langle k | \hat A | k+1\rangle\big). 
$
The quality of the TPQ-MPS is tested by the comparison of $\langle A\rangle_{\beta,N}$ 
with the exact or nearly exact solution by the counterparts like the exact diagonalization (ED), 
QMC or quantum transfer matrix (QTM)\cite{klumper93,klumper98} methods\cite{footcomment}. 
%
\begin{figure}
   \centering
   \includegraphics[width=\columnwidth]{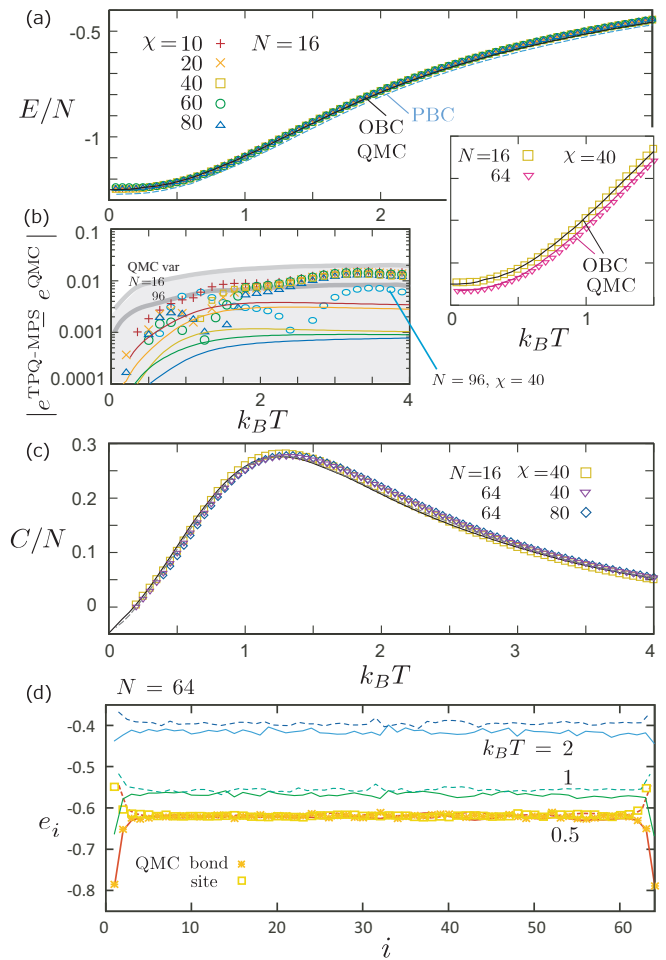}
   \caption{
Comparison of the results of TPQ-MPS and QMC done by authors in the transverse Ising model with $J=g=1$. 
We choose $l=5$ and $k_{max}=500,1200,2000$ for \textcolor{black}{ $N (N_{\it ran})=16 (20),64(5),96(5)$ }, respectively, giving $\beta^{-1}_{max} \sim 0.1$,  
(a) Energy density $e=E/N$ for $N=16$ with $\chi=5,10,20$ and $l=5$ as a function of $k_BT$ 
 compared with the PBC and OBC results from QMC of the same size. 
 Inset shows the low temperature part with $(N,\chi)=(16,40)$ and $(64,40)$ with $l=10$ and the corresponding QMC. 
(b) Energy difference $|e^{\rm TPQ-MPS} - e^{\rm QMC}|$ from (a) and $N=96, \chi=40$, 
  the variance of TPQ-MPS average (solid lines) for $N=16$, $\chi=20-40$ 
  \textcolor{black}{and the variance of the QMC 
  (bold lines above the shades) for 100 independent runs}. 
(c) Specific heat $C/N$ for the same data as (a), and $N=64,96$, $\chi=40$ with $l=5$ in the inset. 
 QMC OBC results for the same sizes are given in solid lines. 
(d)  Spatial distribution of the site- and bond-energies (solid and broken lines) for 
 $N = 64$ and $k_BT = 0.5,1,2$. 
 QMC-OBC results are shown by symbols for $k_BT=0.5$.
}
\label{fig2}
\end{figure}
%
\begin{figure}[tbp]
   \centering
   \includegraphics[width=\columnwidth]{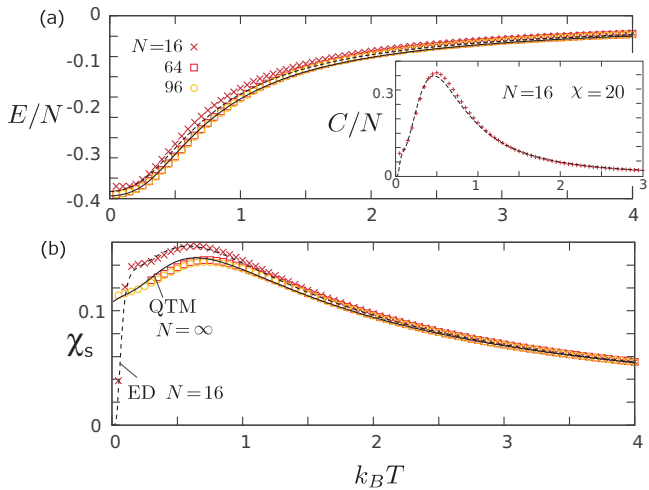}
   \caption{
 Physical properties of the Heisenberg model with $J=1$ compared with 
 the $N=16$ ED-OBC and QTM($N=\infty$ exact solution) ones\cite{klumper98} in broken and solid lines, respectively. 
 We choose $l=1$ and $\beta^{-1}_{max} \sim 0.03$ with $k_{max}=500(N=16), 2000(N=64,96)$. 
(a) Energy density $e=E/N$ and (b) susceptibility $\bm \chi_s$ for $N=16, 64, 96$ with $\chi=20,40,40$ as a function of $k_BT$. 
 The inset in (a) shows the specific heat for $N=16$. 
}
\label{fig3}
\end{figure}
\par
\textcolor{black}{
The required number of random averages of mTPQ runs $N_{\it ran}$ is as small as 
in the full-TPQ that uses the full Hilbert space, e.g. typically less than 5 for $N \gtrsim 32$. 
This can be seen from a benchmark results in Fig.~\ref{fig1}(b); 
When performing 10-mTPQ runs with and without auxiliaries, 
the variance of the former turned out to be small by orders of magnitude, particularly at higher temperature. 
The variance of TPQ-MPS degreases at lower $T$, contrarily to the full-TPQ. 
The higher performance despite its being an approximate method is possibly because 
the basis in the MPS construction is optimally biased at low temperature to those favoring the low energy states. 
}
The difference from the original TPQ also lies that the range of mTPQ temperature $k_BT^{(k)}$ depends much on $l$. 
Usually, starting from smaller $l$ will accelerate the convergence 
and we reach the same temperature with smaller $k_{max}$. 
However, because of finite $\chi$, the starting temperature at $k \sim 1$ for $l \ll \Theta(h)$ 
is kept to $k_BT \lesssim \Theta(h)$ ($h$ being the typical local energy scale of the Hamiltonian) 
in which case the canonical summation becomes inaccurate. 
Whereas, $l\gtrsim \Theta(10h)$ sacrifices the low temperature information, 
and one needs to set proper $l$ depending on the models. 
\\
\textit{Volume law and the auxiliaries.}
Let us first examine the basic entanglement properties of the TPQ-MPS. 
Here, we benchmark the 1D transverse Ising chain, 
$\Hat{H} = J \sum_{i=1}^{N-1} \Hat{\sigma}_i^z \Hat{\sigma}_{i+1}^z -g \sum_{i=1}^N \Hat{\sigma}_i^x$, 
with $\sigma^z=\pm 1$, whose MPO has $D=3$, and take $J=g=1$. 
Figures~\ref{fig1}(c) and \ref{fig1}(d) show the EE, 
$S_i= -{\rm Tr}(\hat\rho_i \ln \hat\rho_i)$ where 
$\hat\rho_i$ is the density matrix when dividing the system into left and right parts at the $i$-th bond. 
It is known that $S_i$ of the TPQ state follows a Page curve\cite{page93}, 
which increases linearly from the edges that reflect the volume law and saturates toward the maximum at the center. 
The standard MPS {\it without auxiliaries} in Fig.\ref{fig1}(d) indeed follows a page-like curve 
at the edge up to the bond dimension where $S_i$ becomes flat because of the upper bound by the bond dimension $\chi$. 
Here, the lower temperature/larger $k$ requires the smaller $\chi$. 
Whereas in TPQ-MPS, the auxiliaries introduces to the edge a large $S_0$ that depends only on the 
effective bond dimension at the edge, generating a nearly flat $S_i$ throughout the system. 
\par
\textcolor{black}{
However, if we divide the system by setting the subsystem at the center, namely cutting the two bonds 
at equal distances from the center, we find physically meaningful entanglement properties. 
Figure~\ref{fig1}(e) is the corresponding EE, 
$S_n^{c}=-{\rm Tr}(\hat\rho_n \ln \hat\rho_n)$, as a function of the size $n$ of the subsystem. 
One finds that $S_n^{c}$ increases linearly in $n$ until it saturates to the upper bound. 
At $k\gtrsim 600$ where the system reaches the temperature $k_BT^{(k)}\lesssim 0.33$, 
$S_n^{c}$ continues to increase up to the very edges of the system. 
}
\\
\textcolor{black}{
\textit{Entropy.} 
These results indicate that the TPQ-MPS wave function has acquired qualitatively different degrees of freedom 
in its state space, just by attaching the two auxiliaries. 
While the upper bound of the EE, $2\ln \chi$, is only twice as large 
as the case without auxiliaries, the EE continues to go up until that bound since the MPS does not feel the edge. 
This is in sharp contrast to the standard MPS where the entanglement is suppressed toward both edges(Fig.~\ref{fig1}(e), bold dots);  
}
\textcolor{black}{
although the bound of $S_n^c$ for the latter was theoretically believed to be $\ln\chi\sim 3.7$(see shaded region), 
the realized value is much smaller $S_n^c \lesssim 1$-$1.5$, and behaves linearly only up to $n\sim 10$ near the edges. 
Because of this advantage, 
}
\textcolor{black}{
the present scheme allows the evaluation of the thermodynamic entropy density $s_{th}$ 
using Eq.(\ref{equality}); 
We perform a linear fit of a series of $S_n^{c}$ for $N=64$ and for various $\chi$, 
and compare its slope with $s_{th}$ obtained by integrating the specific heat of the QMC calculation. 
As shown in Fig.\ref{fig1}(f), the slope of $S_n^{c}$ agrees with $s_{th}$, 
asymptotically approaching the QMC data with increasing $\chi$. 
}
\textcolor{black}{
   In principle, we need to take $\chi \sim {\it e}^{s_{{\it th}} n/2}$ to attain a volume law of 
the subsystem size-$n$. This fact is unrelated to size $N$. 
}
\textcolor{black}{
The TPQ-MPS makes full use of the theoretical bound of $S_n\lesssim 2\ln \chi=4-10$  for $\chi=50-100$ 
which affords the description of the entanglement of the pure state at the standard target temperature range, 
e.g. $k_BT \le J$, that are of physical interest in typical quantum lattice models. 
}
\textcolor{black}{
By contrast, $S_n$ of the usual MPS is unreliable because it is tightly bounded by the peak of 
the Page curve related to $N$, which is lower than $\ln \chi$. 
}
\\
\textit{Benchmarks.} 
Figure \ref{fig2}(a) shows the energy density $e=E/N$ of the transverse Ising model 
for $N=16$ which is in good agreement with the QMC results 
with open boundary condition (OBC), where we put together the periodic boundary (PBC) ones of the same size. 
The $N=16$ and 64 cases are also compared in the inset, 
demonstrating that the finite size effect is much larger than the difference between the TPQ-MPS and QMC results. 
Already at $\chi \gtrsim 10$, the error is converged as we see in Fig.\ref{fig2}(b) 
which is smaller than \textcolor{black}{the variance of the QMC results over 100 independent runs each with 200000 averages}, 
regardless of size-$N$ and $\chi$. 
The specific heat $C/N$ also gives excellent agreement with the QMC results of the same size
(see Fig.\ref{fig2}(c)). 
Figure \ref{fig2}(d) gives the spatial distribution of the bond- and site-energies 
($J$- and $g$-terms of the Hamiltonian) for several temperatures at $N=64$. 
They perfectly follow those of OBC obtained by QMC for $k_BT=0.5$ even at the very edges that show downturn/upturn. 
The variance of mTPQ runs are kept small enough; 10-20 averages for $N=16$ and 5 for $N=64,96$ in the figures. 
\par
Next, we test our method with the Heisenberg chain described by 
$\Hat{H} =\sum_{i=1}^{N-1} J  \Hat{s}_i \Hat{s}_{i+1}$, with $s_i^z=\pm 1/2$. 
The MPO for this Hamiltonian has $D=5$ that would increase the truncation error. 
In fact, it is known for the MPS ground state that the model requires much larger $\chi$ 
compared to the product-type ground state of the transverse Ising model. 
Figure \ref{fig3}(a) shows $E/N$ and $C/N$ for several system sizes in comparison with the 
$N=16$ ED with OBC and $N=\infty$ exact solution (QTM)\cite{klumper98}, 
which shows good agreement for the same order of $\chi$ as the transverse Ising model. 
We also plot in Fig. \ref{fig3}(b) the susceptibility $\chi_s$ to see how much a well-known logarithmic singularity 
at the lowest temperature can be traced by the TPQ-MPS. 
The drop of $\chi_s$ at $N=16$ is almost perfectly reproduced already for $\chi=20$. 
Also, the larger size results are in reasonable agreement with QTM. 
\par
\textit{Conclusion.} 
\textcolor{black}{We realized the TPQ-MPS by attaching the edge-auxiliaries of dimension $\chi$ to the MPS, 
showing that the EE of the subsystem cut out from the center 
continues to show a volume law up to the very edges of the system, particularly at low temperature, 
by setting the bond dimension $\chi$ to the realistically small values. 
Physical properties are evaluated almost free of random sampling. 
}
Such construction enables the application to higher dimensions, possibly less costly than the aforedeveloped 
tensor network approaches\cite{chen18,han19,bruognolo17,xiang19}. 
\par
\textit{Acknowledgements.} We thank Tsuyoshi Okubo and Hal Tasaki for discussions. 
The work is supported by JSPS KAKENHI Grants No. JP17K05533, No. JP18H01173,
No. JP17K05497, No. JP17H02916, and No.JP19H01810. 

\bibliography{tpqmps}

\begin{thebibliography}{51}
\bibitem [{foo({\natexlab{a}})}]{footcomment2}%
  \BibitemOpen
  \href@noop {} {} ({\natexlab{a}}),\ \bibinfo {note} {in the mTPQ algorithm,
  $k_BT^{(k)} = N(l-u_k)/2k$ gives a rough estimate of the microcanonical
  temperature of $\Theta(1/N)$. Higher order corrections can be evaluated
  following Ref.\cite{sugiura12}.}\BibitemShut {Stop}%
\bibitem [{foo({\natexlab{b}})}]{footcomment3}%
  \BibitemOpen
  \href@noop {} {} ({\natexlab{b}}),\ \bibinfo {note} {we define the truncation
  error as the weight discarded in the Schmidt decomposition. In the transverse
  Ising model, the maximum truncation error at each $k$ is typically
  $10^{-4}-10^{-5}$ for $k_BT^{(k)}\lesssim 1$ at $N=16$ and decreases by
  orders of magnitude with both $N$ and $\chi$.}\BibitemShut {Stop}%
\bibitem [{foo({\natexlab{c}})}]{footcomment}%
  \BibitemOpen
  \href@noop {} {} ({\natexlab{c}}),\ \bibinfo {note} {in the previous TPQ
  formalism where $\ket{\psi^{\rm RM}}$ in Eq.(\ref{mtpq}) is the random state
  constructed using the full Hilbert space, the variance of $\langle \Hat{A}
  \rangle ^{\text{TPQ}} _k$ against the ensemble average $\langle \Hat{A}
  \rangle = \text{Tr} [ \Hat{\rho}_k \Hat{A} ]$, is proved to be as small as of
  the order $e^{-\Theta(N)}$\cite{sugiura12,sugiura13}. Although it is no
  longer guaranteed, the following demonstration shows comparably accurate
  results for the TPQ-MPS.}\BibitemShut {Stop}%
\end{thebibliography}
\end{document}